\documentclass[Preprint,12pt]{elsarticle}
\usepackage{newtxtext}

\usepackage{lineno}

\usepackage[bottom]{footmisc} 

\usepackage{geometry}
\usepackage{amsmath}
\usepackage{algorithm}
\usepackage{algpseudocode}
\usepackage{url}
\usepackage{ upgreek }
\usepackage{pdflscape}
\usepackage{caption}
\usepackage{setspace}
\usepackage{pifont}
\usepackage{amsmath}
\usepackage{bbding}
\usepackage{amsfonts} 
\usepackage{amssymb}
\usepackage{mathrsfs}

\usepackage{color}
\usepackage{ dsfont }
\usepackage[misc]{ifsym}

\usepackage{amsmath}      
\usepackage{graphicx}  
\usepackage{array}
\usepackage{verbatim}
\usepackage{titletoc}
\usepackage{color,xcolor}

\usepackage{units}
\usepackage{multirow}
\usepackage{tabularx}
\usepackage{booktabs}
\usepackage{bbm}
\usepackage{amsmath}      
\usepackage{bm} 
\usepackage{titlesec}
\usepackage{algorithm}
\usepackage{algpseudocode}
\titleformat{\paragraph}
  {\normalfont\normalsize\itshape}{\theparagraph}{1em}{}
\titlespacing*{\paragraph}
  {0pt}{1.5ex plus 0.5ex minus .2ex}{0.5ex plus .2ex}  

\usepackage{amsmath}

\usepackage{xcolor}
\definecolor{DeBlue}{RGB}{0,0,0}
\definecolor{DeGreen}{RGB}{84,130,53}
\definecolor{DeBlack}{RGB}{0,0,0}

\usepackage{natbib}
\usepackage{fancyhdr}

\begin{document}

\title{IDS-Net: A novel framework for few-shot photovoltaic power prediction with interpretable dynamic selection and feature information fusion}
\begin{frontmatter}

\author[inst1,inst2]{Hang Fan}
\affiliation[inst1]{organization={School of Economic and Management},
            addressline={North China Electric Power University}, 
            city={Changping District, Beijing, 102206},
            country={China}}

\affiliation[inst2]{organization={Beijing Key Laboratory of Renewable Energy and Low-carbon Development},
            addressline={Changping District}, 
            city={Beijing, 102206},
            country={China}}

\affiliation[inst3]{organization={School of Electrical and Electronic Engineering},
             addressline={Nanyang Technological University}, 
             city={50 Nanyang Avenue, 639798},
             country={Singapore}}

\author[inst3]{Weican Liu\corref{cor1}}\footnotesize
\cortext[cor1]{W. Liu is the corresponding author (Email: weican001@e.ntu.edu.sg).}

\author[inst1]{Zuhan Zhang} 
\author[inst1]{Ying Lu}

\author[inst1]{Wencai Run}

\author[inst1,inst2]{Dunnan Liu} 
\normalsize
\begin{abstract}
With the growing demand for renewable energy, countries are accelerating the construction of photovoltaic (PV) power stations. However, accurately forecasting power data for newly constructed PV stations is extremely challenging due to limited data availability. To this end, we propose a novel interpretable dynamic selection network (IDS-Net) based on feature information fusion to achieve accurate few-shot prediction. This transfer learning framework primarily consists of two parts. In the first stage, we pre-train on the large dataset, utilizing Maximum Mean Discrepancy (MMD) to select the source domain dataset most similar to the target domain data distribution. Subsequently, the ReliefF algorithm is utilized for feature selection, reducing the influence of feature redundancy. Then, the Hampel Identifier (HI) is used for training dataset outlier correction. In the IDS-Net model, we first obtain the initial extracted features from a pool of predictive models. Following this, two separate weighting channels are utilized to determine the interpretable weights for each sub-model and the adaptive selection outcomes, respectively. Subsequently, the extracted feature results from each sub-model are multiplied by their corresponding weights and then summed to obtain the weighted extracted features. Then, we perform cross-embedding on the additional features and fuse them with the extracted weighted features. This fused information is then passed through the MLP (Multi-Layer Perceptron) layer to obtain predictions. In the second stage, we design an end-to-end adaptive transfer learning strategy to obtain the final prediction results on the target dataset. We validate the transfer learning process using two PV power datasets from Hebei province, China, to demonstrate the effectiveness and generalization of our framework and transfer learning strategy.
\end{abstract}


\end{frontmatter}
\setcounter{page}{1}

\section{Introduction}

\subsection{Background and motivation}

As economic development progresses, global energy demand continuously increases \cite{dai2025short}. Among these energies, solar energy has garnered widespread attention due to its inexhaustible, renewable, and green energy characteristics \cite{peng2024short}. By 2023, China's solar PV installed capacity accounted for 16.6\% of the world's total \cite{LI2025122991}. However, PV power data is characterized by intermittency and volatility, posing numerous challenges for accurate PV forecasting \cite{wang2025cross}. These challenges are particularly pronounced for newly constructed PV power stations due to the lack of sufficient historical data for model training, which makes it difficult to achieve precise prediction results. Therefore, developing an efficient and accurate transfer strategy to address the few-shot problem in PV forecasting is particularly crucial \cite{liao2024enhanced}.

\vspace{0.25\baselineskip}

In terms of deep learning architectures, enabling models to directly acquire data distribution representations from historical data is undeniably an efficient approach. Nevertheless, for newly constructed photovoltaic power stations, the scarcity of historical data—often limited to merely three to five days—frequently renders it difficult for models to extract meaningful information. Furthermore, despite the ability of various existing models to achieve superior prediction accuracy, they all possess certain limitations. For example, although current PV forecasting studies incorporate weather variables as features, they often overlook the complex and dynamic interaction between weather conditions and PV power generation. Moreover, existing research frequently employs only a single model for PV power forecasting, which significantly compromises the model's prediction performance across different datasets. Even when ensemble learning techniques are utilized, they fail to simultaneously account for both interpretability and dynamic adaptive selection. To this end, we propose a novel PV forecasting paradigm, which includes a rapid transfer learning strategy, a parallel dual-channel ensemble strategy, and a multi-feature interaction collaborative learning strategy, designed to solve the few-shot prediction problem in PV power stations.

\subsection{Related work}
Accurate PV power forecasting is crucial for improved energy planning and power transmission management. It helps mitigate the impact of PV power volatility on grid stability. With the continuous advancement of various forecasting techniques, PV power generation prediction models can primarily be categorized into three types: physical models, statistical models, and artificial intelligence (AI) models \cite{liu2025pv}. Physical models do not require historical data for model training. Instead, they predict PV power generation primarily based on the dynamic interaction between the solar radiation in the atmosphere and the key design parameters of the PV system \cite{tian2025enhancing}. For example, Ogliari \textit{et al.} proposed a physical PV forecasting model that utilizes weather conditions. This model achieved superior prediction accuracy by deriving the solar-to-electrical energy conversion process from the intrinsic parameters of the photovoltaic panel \cite{ogliari2017physical}. Despite their robust theoretical basis and strong interpretability, physical models face limitations due to the intricate nature of atmospheric phenomena, which leads to increased model complexity \cite{tian2025pvmtf}. In contrast, statistical models leverage rigorous mathematical derivations for future PV power forecasting. Common examples include Autoregressive Moving Average (ARMA) and Autoregressive Integrated Moving Average (ARIMA) \cite{phinikarides2013arima}. These models utilize historical data as input to statistically derive the likely PV power. For example, Li \textit{et al.} improved the traditional ARIMA model by incorporating meteorological features such as temperature, precipitation, and sunshine duration as exogenous inputs. This approach not only retained the simplicity of the ARIMA model but also further enhanced PV prediction accuracy \cite{li2014armax}. However, statistical models are typically based on various mathematical assumptions to obtain predictions, making it challenging to capture the nonlinear relationships among multivariate time series data \cite{liu2025enhancing}. With the continuous advancement of computer technology, the emergence of artificial intelligence has propelled the development of deep learning techniques, which are gradually becoming the mainstream models for PV forecasting \cite{luan2025ensemble}. Some fundamental deep learning models, such as CNN \cite{agga2021short}, RNN \cite{vu2022optimal}, and LSTM \cite{riedel2024enhancing}, have been widely applied in the field of PV forecasting, demonstrating the developmental potential of AI-driven models. To further enhance PV prediction accuracy, researchers have begun to combine different deep learning models, aiming to overcome the limitations of single prediction models \cite{tian2025new}. For example, Abou \textit{et al.} proposed the COA-CNN-LSTM model for PV forecasting, which utilizes the coati optimization algorithm to optimize the hyperparameters of CNN-LSTM, achieving the optimum parameters and performance compared to other models \cite{abou2023coa}. Bashir \textit{et al.} proposed two hybrid models: CNN-ABiLSTM and CNN-transformer-MLP model for PV and wind forecasting, achieving the best predictions compared to other rivals \cite{bashir2025wind}. Subsequently, with the aim of enhancing forecasting accuracy, hybrid models became extensively utilized in PV prediction. Various components that enhance prediction accuracy have been applied in the field of PV forecasting, such as feature engineering, denoising data techniques, and error correction strategies \cite{tian2025edformer}. For example, Li \textit{et al.} proposed a hybrid medium-term PV forecasting model based on the ranked feature selection and genetic algorithm \cite{li2025wnps}. Zhu \textit{et al.} proposed a novel hybrid probabilistic PV forecasting model based on evolutive chaotic particle swarm optimization, improving the accuracy and reducing the computational burden \cite{zhu2025novel}. Chen \textit{et al.} proposed a novel PV forecasting model, CGAformer, which utilizes CNN blocks and multi-head GADAttention blocks to select and extract multi-scale features. Finally, the forecasting output is obtained through an MLP, effectively improving PV prediction accuracy across different seasons \cite{chen2024cgaformer}.

\vspace{0.25\baselineskip}
In recent years, the ensemble learning prediction paradigm has seen extensive application in PV forecasting. As a type of hybrid model, its widespread use is primarily attributed to its notable robustness and generalization capabilities \cite{tian2024cnns}. For example, Atiea \textit{et al.} proposed a PV forecasting model based on Bayesian optimization and stacked ensemble learning, improving the accuracy of PV generation \cite{atiea2025photovoltaic}. Xu \textit{et al.} proposed a double-layer PV forecasting model, based on Blending ensemble learning considering different weather types, which achieves a higher fitting accuracy \cite{xu2025double}. Despite the promising prediction accuracy achieved by the ensemble learning prediction paradigm in the field of PV forecasting, there remains room for improvement. Currently, mainstream ensemble learning forecasting methods can primarily be categorized into two main types. The first type, called the parallel weighted ensemble learning method, involves each sub-model making parallel predictions. The optimal sub-model prediction results are then selected and weighted based on specific prediction metrics \cite{xu2024paradigm}. However, this often presents the challenge of objective function mismatch. This is because a higher prediction accuracy in an individual sub-model does not guarantee that the selected high-accuracy sub-models will achieve the highest precision after weighting. To address this challenge, researchers proposed a second type, named the sequential ensemble learning method \cite{WANG2024130305, WANG2025134884}. Specifically, this involves multi-layer model learning integration, where sub-models first generate prediction results, and these predictions are then used as feature input into the next layer's learner to obtain the final prediction output. Although the second method utilizes an end-to-end approach, resolving the objective function mismatch problem, it struggles to provide effective interpretability. This makes it difficult to establish a clear causal relationship between the sub-model predictions and the final forecasting output. To this end, this paper proposes a dual-channel ensemble learning PV forecasting model that achieves both interpretable weights and adaptive selection.

\vspace{0.25\baselineskip}
Furthermore, traditional PV models are typically trained on large volumes of historical data. Consequently, they often struggle to achieve optimal prediction accuracy when faced with newly constructed power stations that lack sufficient historical data. In recent years, some scholars have started to focus on the PV few-shot prediction problem, utilizing transfer learning strategies to improve prediction accuracy. For example, Xu \textit{et al.} proposed a Swin Transformer model based on transfer learning to improve few-shot PV forecasting accuracy, which demonstrates the feasibility of transfer learning in addressing the few-shot prediction problem \cite{xu2025swin}. However, this research assumes that fine-tuning on the target dataset is necessary. Since the fine-tuning process still requires freezing model parameters and subsequent training, it further increases computational complexity. Therefore, our proposed rapid transfer strategy can automatically determine whether fine-tuning is necessary for results, effectively reducing computational burden while ensuring prediction accuracy. \textbf{Table 1} contrasts our work with the literature. To the best of our knowledge, this work represents the first instance of combining interpretable dynamic adaptive selection ensemble learning with an end-to-end rapid transfer strategy for PV forecasting. This significantly enhances few-shot prediction accuracy while simultaneously providing rich interpretability.

\begin{table}[h]
\small
\begin{spacing}{1}
\caption{\textbf{Contrasting our work with the literature}}
\setlength{\tabcolsep}{2.8mm}{
\begin{tabular}{ccccccc}
\hline
\multicolumn{2}{c}{\multirow{3}{*}{Reference}} & \multirow{3}{*}{\begin{tabular}[c]{@{}c@{}}Data \\ Preprocessing\end{tabular}} & \multirow{3}{*}{\begin{tabular}[c]{@{}c@{}}Ensemble\\ Learning\end{tabular}} & \multirow{3}{*}{\begin{tabular}[c]{@{}c@{}}Interpretable \\ Selection\end{tabular}} & \multirow{3}{*}{\begin{tabular}[c]{@{}c@{}}Transfer\\ Learning \end{tabular}} & \multirow{3}{*}{\begin{tabular}[c]{@{}c@{}}Information\\ Fusion\end{tabular}} \\
\multicolumn{2}{c}{}       &      &       &      &      &    \\
\multicolumn{2}{c}{}       &      &       &      &      &    \\ \hline
\textit{\begin{tabular}[c]{@{}c@{}} Physical \& \\ Statistical \\ Model\end{tabular}} & \cite{ogliari2017physical, phinikarides2013arima, li2014armax, WANG2022121946} & \ding{55}    & \ding{55}  & \ding{55}      & \ding{55}               & \ding{55}        \\
\multirow{4}{*}{\textit{AI Based}} &  \cite{abou2023coa, bashir2025wind, li2025wnps, zhu2025novel, chen2024cgaformer}    & \checkmark         & \ding{55}              & \ding{55}      & \ding{55}       & \checkmark   \\
 &  \cite{atiea2025photovoltaic, xu2025double, xu2024paradigm}       & \checkmark         & \checkmark              & \ding{55}        & \ding{55}       & \checkmark  \\
   & \cite{xu2025swin}        & \checkmark         & \ding{55}               & \ding{55}        & \checkmark       & \checkmark  \\
 & \textbf{Our Works} & \textbf{\checkmark}  &\textbf{\checkmark}    & \textbf{\checkmark}       & \textbf{\checkmark}  & \textbf{\checkmark}        \\ \hline
\end{tabular}}
\end{spacing}
\end{table}

\subsection{Our contributions}

To effectively address the limited historical data issue in newly constructed PV stations and to provide rich interpretability for prediction, we propose a novel few-shot PV prediction framework, namely IDS-Net. Specifically, our framework is mainly divided into two parts: a data preprocessing module and a PV power prediction module. In the data preprocessing module, we first utilize the MMD to select the pretraining dataset. Subsequently, we perform outlier correction on the training set and reduce feature redundancy using feature engineering. In the PV prediction module, we propose a dual-channel prediction module, comprising an interpretable weighting channel and an adaptive selection channel, to simultaneously enhance the interpretability and selectivity of the ensemble learning. Finally, after independent feature learning of additional PV features, the extracted PV features and PV prediction features are cross-fused, and the final prediction results are obtained through an MLP. \textbf{Thus, the major contributions of this paper can be summarized as follows:} 

\vspace{-0.5\baselineskip}

\begin{itemize}
\item 
\emph{\textbf{An accurate few-shot PV forecasting framework based on transfer learning}}: To address the limited historical data problem of newly constructed  PV stations, we design a transfer learning PV prediction framework, primarily divided into three parts: preprocessing, prediction, and transfer. First, we utilize MMD to select a pretraining dataset and perform data preprocessing on the training set. We then construct the prediction ensemble framework. Subsequently, the pre-trained model can be transferred to the target domain through fine-tuning. Experiments show that our proposed fast transfer strategy effectively improves the prediction accuracy.

\item 
\emph{\textbf{The effective data preprocessing module for outlier correction and feature selection}}: PV power data is characterized by complexity and uncertainty and is also significantly influenced by weather factors. To reduce the impact of outliers and feature redundancy in the training data on model training, our data preprocessing module is primarily divided into three parts: data selection, outlier correction, and feature engineering (MMD-HI-ReliefF). First, we use the MMD algorithm to select an appropriate pre-training dataset. Then, we apply the HI algorithm for outlier correction to improve the quality of the training data. Finally, we utilize the ReliefF algorithm to select relevant features and reduce feature redundancy.

\item 
\emph{\textbf{The interpretable dynamic selection prediction module based on information fusion}}: To enhance the interpretability of prediction models and fully leverage weather feature information, we propose a dual-channel interpretable dynamic selection prediction model. The first channel constructs a learnable weighting function, where a larger weight indicates a greater contribution from that sub-model. The second channel builds a selection function, which represents whether to select or discard the given sub-model. Finally, the use of additional weather features information for corrective fusion further improves the accuracy of PV forecasting.

\end{itemize}

The rest of the paper proceeds as follows: Section 2 introduces the problem description and framework design. Section 3 introduces the data preprocessing module, the design of the IDS-Net structure, the customized loss function and the effective transfer learning strategy. Section 4 introduces data selection, parameter design, evaluation indexes, and the operation environments. Section 5 demonstrates the performance of the proposed few-shot PV forecasting framework compared with the baseline models through four experiments. Finally, Section 6 introduces the conclusion and future works.

\section{The Problem Description and Framework Design}

This section provides a detailed exposition of the problem our research tackles, coupled with a comprehensive explanation of our proposed few-shot PV forecasting framework.

\subsection{The problem description of few-shot PV forecasting}

In this section, we introduce the preliminaries of the few-shot PV forecasting problem. In this paper, the prediction task is to forecast the future PV generation of newly constructed PV stations, which have limited historical data and are referred to as target domain data. During the pre-training stage, we utilize data from mature PV stations (with sufficient PV power generation data) that had the most similar data distribution to the target domain data. These are referred to as source domain data.

\vspace{0.25\baselineskip}

\textit{\textbf{Definition 1. PV Data}}: We define the PV data as $\mathbf{X} \in \mathbb{R}^{BS \times LW \times F}$, where $BS$ is the Batch Size, $LW$ is the Look-back Windows, and $F$ is the number of features. For PV power data, the $F$ is 1; for additional weather data features related to PV power generation, $F$ is defined as $N$.

\vspace{0.25\baselineskip}

\textit{\textbf{Definition 2. PV Prediction}}: Given the input PV power data is $\mathbf{X}_P = \{\mathbf{x}_{t-P+1}, \mathbf{x}_{t-P+2}, \ldots, \mathbf{x}_t\} \in \mathbb{R}^{BS \times P \times 1}$, where we define $LW$ is $P$ and $F$ is 1. And features are $\mathbf{F}_P = \{\mathbf{F}_{t-P+1}, \mathbf{F}_{t-P+2}, \ldots, \mathbf{F}_t\} \in \mathbb{R}^{BS \times P \times N}$, where we define $LW$ is $P$ and $F$ is $N$. So the objective is to train a function $f(\cdot)$ with parameter $\theta$ to predict the future $S$ time steps, $\mathbf{Y}_s = \{\mathbf{y}_{t+1}, \mathbf{y}_{t+2}, \ldots, \mathbf{y}_{t+s}\} \in \mathbb{R}^{BS \times P \times S}$. In this work, we define $S$ as 1. The formula of prediction progress can be defined as follows:

\vspace{-0.5\baselineskip}
\begin{equation}
\mathbf{Y}_s = f(\mathbf{X}_P, \mathbf{F}_P; \theta).
\end{equation}

\textit{\textbf{Definition 3. PV Few-shot Learning Prediction}}: Assume that the few-shot PV data and weather features data are as $\mathbf{X}_{pt}$ and $\mathbf{F}_{pt}$, which denotes the target domain. $\mathbf{X}_P$ and $\mathbf{F}_P$ denotes the source domain. The data numbers in the target domain are significantly less than those in the source domain. And the  $T(\cdot)$ denotes the rapid transfer learning function. The objective of few-shot learning prediction is to utilize the pre-training model using the source domain to learn a predictive function, which can be defined as follows:

\vspace{-0.5\baselineskip}
\begin{equation}
\mathbf{Y}_{st} = T(\mathbf{X}_{pt}, \mathbf{F}_{pt}; f(\mathbf{X}_P, \mathbf{F}_P; \theta)),
\end{equation}
where $\mathbf{Y}_{st}$ is the prediction results in the target domain.

\subsection{The design of the whole framework} 

\begin{figure}[!b]
    \centering
    \captionsetup{labelfont=bf}
    \includegraphics[width=0.96\linewidth]{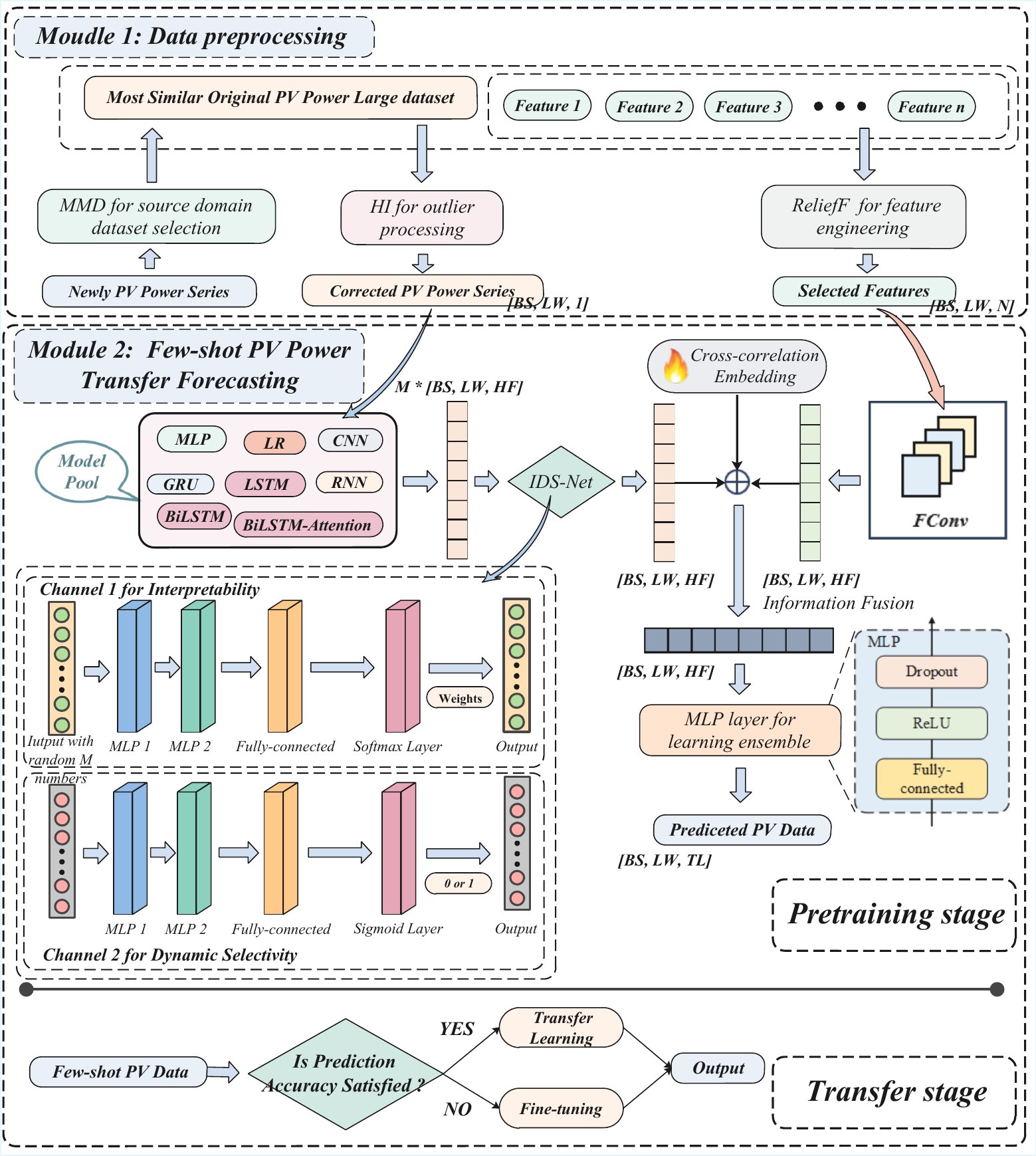}
    \vspace{-0.4\baselineskip}
    \caption{\textbf{Flow chart of IDS-Net framework}}
\end{figure} 

This study proposes a transfer learning PV power forecasting framework based on interpretable dynamic selection and multi-feature information fusion to address the issue of insufficient historical data for newly constructed PV power stations. Our research is divided into two modules: the data preprocessing module and the few-shot PV power forecasting module. In the data preprocessing module, we first apply the HI to the training set of PV data to handle outliers, reducing the impact of anomalous noise. Subsequently, we utilize the ReliefF to select relevant features, lowering the complexity of model training. In the IDS-Net forecasting module, we first obtain multiple sets of prediction results using a deep learning model pool. Subsequently, a dual-channel neural network is employed to enhance both the interpretability and dynamic selectivity of the sub-models. Within the information fusion block, we fuse the sub-model prediction results with additional feature information, and finally, an MLP layer is utilized for the final prediction. \textbf{Figure 1} demonstrates the flow chart of our proposed IDS-Net framework. 

\section{Methodology}

In this section, we first introduce the data preprocessing modules, including the MMD, HI, and ReliefF algorithms. Next, we introduce the proposed IDS-Net, including the deep learning model pool, the two dual-channel structure, and the information fusion strategy. Finally, we introduce the transfer learning strategy and the customized loss function mechanism in detail.

\subsection{Effective data preprocessing module}

\begin{figure}[!b]
    \centering
    \captionsetup{labelfont=bf}
    \includegraphics[width=0.96\linewidth]{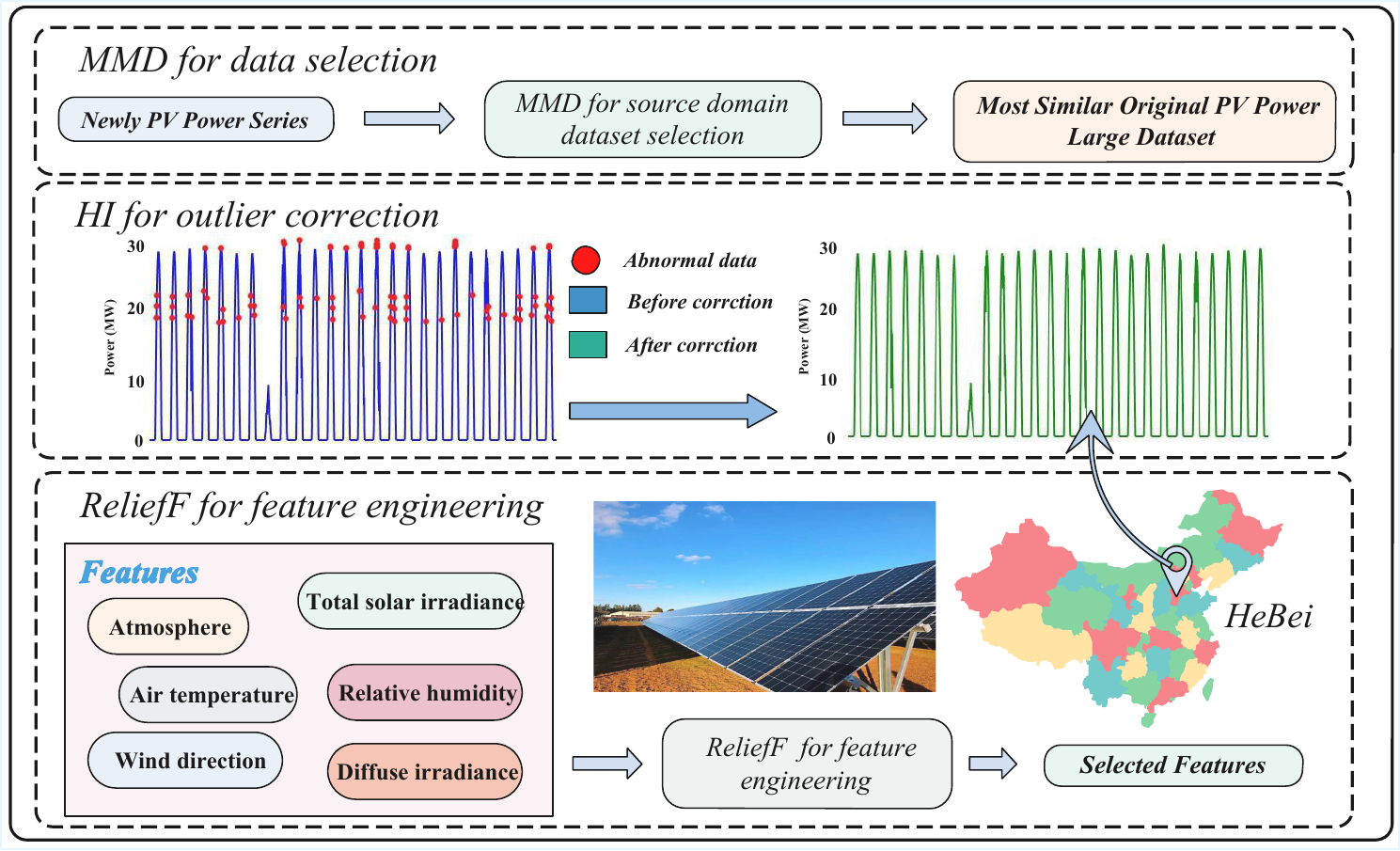}
    \vspace{-0.2\baselineskip}
    \caption{\textbf{Flow chart of the effective data preprocessing module}}
\end{figure} 

Considering that PV data are significantly influenced by meteorological factors, exhibiting complexity and high volatility, effective data preprocessing is crucial for greatly improving PV prediction accuracy. First, our task addresses the challenge of achieving high-precision prediction results for newly constructed PV power stations that lack sufficient historical data. Therefore, we utilize the MMD technique to select the power station with the most similar data distribution from adjacent PV power stations with ample training data, serving as the source domain for pre-training. Next, the HI algorithm is applied to the source domain training set for outlier correction, reducing the impact of anomalies on model training. Finally, the ReliefF algorithm is employed to select the most relevant feature variables, mitigating the influence of feature redundancy on model training. The data preprocessing module is shown in \textbf{Figure 2}.

\subsubsection{Maximum Mean Discrepancy for data selection}

In few-shot PV forecasting, selecting the appropriate pre-training dataset is critical, as it directly determines the prediction accuracy after transfer learning \cite{fan2025privacy}. Considering that PV power plants located in the same geographical area as a newly constructed PV station are likely to exhibit similar power distribution patterns, we utilize Maximum Mean Discrepancy (MMD) to select the station with the most similar distribution patterns within that region. We utilized a multi-kernel variant of MMD, which maps data distributions into a Reproducing Kernel Hilbert Space (RKHS). The squared MMD between two distributions is denoted as $d_1$ and $d_2$, respectively. The formulation of MMD can be defined as follows:

\begin{equation}
    \text{MMD}^2 (d_1, d_2) = \| E[\phi(d_1)] - E[\phi(d_2)] \|,
\end{equation}
where $\phi(\cdot)$ denotes the mapping to RKHS. In this work, we utilize the kernel function to replace this mapping process. Therefore, the MMD can be defined as follows:

\begin{equation}
    \text{MMD}^2 (d_1, d_2) = E[K(d_1, d_1)] + E[K(d_2, d_2)] - 2E[K(d_1, d_2)],
\end{equation}
where $K(d_1, d_1) = [\phi(x), \phi(y)]$ denotes the kernel function, which is $\exp\left(-\frac{\|\mathbf{d_1} - \mathbf{d_2}\|^2}{2\sigma^2}\right)$. Therefore, we select the dataset with the minimum MMD value as the source domain for the pre-training stage.

\subsubsection{Hampel Identifier for outlier correction}

Traditional outlier detection methods predominantly rely on the "three-sigma rule," which is based on the mean and standard deviation. However, these methods are easily perturbed by the outliers themselves \cite{yang2025hybrid}. To overcome this issue, we utilize the HI algorithm based on the median and the Median Absolute Deviation (MAD). The formula can be defined as follows:

\begin{equation}
    m_i = \text{median} (x_{i-k}, x_{i-k+1}, x_{i-k+2}, \ldots, x_i, \ldots, x_{i+k-2}, x_{i+k-1}, x_{i+k}),
\end{equation}

\begin{equation}
    MAD_i = \text{median} (|x_{i-k} - m_i|, \ldots, |x_{i+k} - m_i|),
\end{equation}

\begin{equation}
    \sigma_i = \lambda \times MAD_i,
\end{equation}

\begin{equation}
    |x_i - m_i| > n_{\sigma} \times \sigma_i,
\end{equation}
where $x_{i}$ denotes the data in the dataset, $k$ denotes the length of the sliding window, $m_i$ denotes the median value of the sliding window. $MAD_i$ represents the median of the absolute differences between each data point and the local median within the same sliding window. $\sigma_i$ is the standard deviation, $\lambda$ can be defined as 1.4826. $n_{\sigma}$ is the 
where, $x_{i}$ denotes the data in the dataset, $k$ denotes the length of the sliding window, $m_i$ denotes the median value of the sliding window. $MAD_i$ represents the median of the absolute differences between each data point and the local median within the same sliding window. $\sigma_i$ is the standard deviation, $\lambda$ can be defined as 1.4826. $n_{\sigma}$ denotes a predefined threshold, beyond which a data point is identified as an outlier. Once an outlier is identified, we replace the outlier $x_i$ with the local median ($m_i$).

\subsubsection{ReliefF for feature engineering}

In our work, we employ the ReliefF algorithm to select feature variables most relevant to PV power. ReliefF is an extension of the original Relief algorithm, designed to overcome its limitations in handling multi-class feature identification problems \cite{wang2025relieff}. Specifically, it effectively enhances robustness to noise by computing information from k nearest features, rather than processing only the single nearest feature. The feature \( A \) can be updated weight by the formula as follows:

\begin{equation}
\text{W}(A) = \text{W}(A) - \sum_{j=1}^{k} \frac{\text{diff}(A, R, H_j)}{mk} + \sum_{C \neq \text{class}(R)}  \left[ \frac{\text{P}(C)}{1-\text{P}(\text{class}(R))} \sum_{j=1}^{k} \frac{\text{diff}(A, R, M_{j}^{(C)})}{mk} \right],
\end{equation}
where $W(A)$ denotes the calculated weight for attribute $A$. $R$ refers to an arbitrarily selected data instance. $H_j$ represents the $j$-th closest 'hit' -- specifically, a neighboring instance belonging to the same class. Conversely, $M_{j}^{(C)}$ signifies the $j$-th closest 'miss,' which is a neighboring instance from a distinct class $C$. The parameter $k$ indicates the number of nearest neighbors taken into account during the calculation, and $m$ represents the overall quantity of instances sampled. The operational mechanism of the difference function, $\text{diff}(A, R_1, R_2)$, is described below:

\begin{equation}
\text{diff}(A, R_1, R_2) = 
\begin{cases}
\frac{|R_1[A] - R_2[A]|}{\text{max}(A) - \text{min}(A)} & \text{if } A \text{ is Continuous} \\
0 & \text{if } A \text{ is Discrete and } R_1[A] = R_2[A] \\
1 & \text{if } A \text{ is Discrete and } R_1[A] \neq R_2[A]
\end{cases}.
\end{equation}

\subsection{The design of the IDS-Net structure}

In this section, we will provide a detailed introduction to our proposed deep learning model pool, the dual-channel MLP parameter learning structure, and the multi-feature information interaction and fusion strategy, respectively.

\subsubsection{The deep learning model pool}

In constructing the deep learning model pool, we fully consider the inherent characteristics of PV power data and the current research state of PV forecasting. We meticulously select various deep learning models, each exhibiting particular strengths and proven predictive superiority across different datasets. Within the model pool, we employ fundamental linear models such as Linear Regression (LR) and Multilayer Perceptron (MLP) to quickly capture linear relationships within the data. Additionally, sequential deep learning models, represented by Convolutional Neural Networks (CNNs), Recurrent Neural Networks (RNNs), Gated Recurrent Units (GRUs), Long Short-Term Memory (LSTM), and Bidirectional Long Short-Term Memory (BiLSTM) networks, are utilized to capture the time-varying trends and nonlinear features of PV power data. Furthermore, we also utilize hybrid models, such as the BiLSTM-Attention model, where a BiLSTM combined with a multi-head attention mechanism exhibits stronger sequence modeling capabilities. By integrating multiple models that demonstrate unique advantages in diverse scenarios and datasets, we can more comprehensively address the complexities of PV power forecasting and enhance the prediction accuracy of a single model by leveraging their respective strengths.

\subsubsection{The dual-channel parameter learning structure}

In this section, we introduce a dual-channel parameter learning architecture designed to enhance the interpretability and dynamic adaptive selectivity of ensemble learning models, respectively. The current ensemble learning research can be divided into two parts. First, methods often rely on sub-model accuracy-based selection, which frequently leads to a mismatch between high sub-model accuracy and the ensemble method's objective, making optimal accuracy difficult to achieve. Alternatively, SHAP-based explainable methods provide interpretability, but sub-model selection is often subject to researchers' subjective influence, hindering intelligent implementation. \textbf{Figure 3} demonstrates this detailed structure.

\begin{figure}[!h]
    \centering
    \captionsetup{labelfont=bf}
    \includegraphics[width=0.9\linewidth]{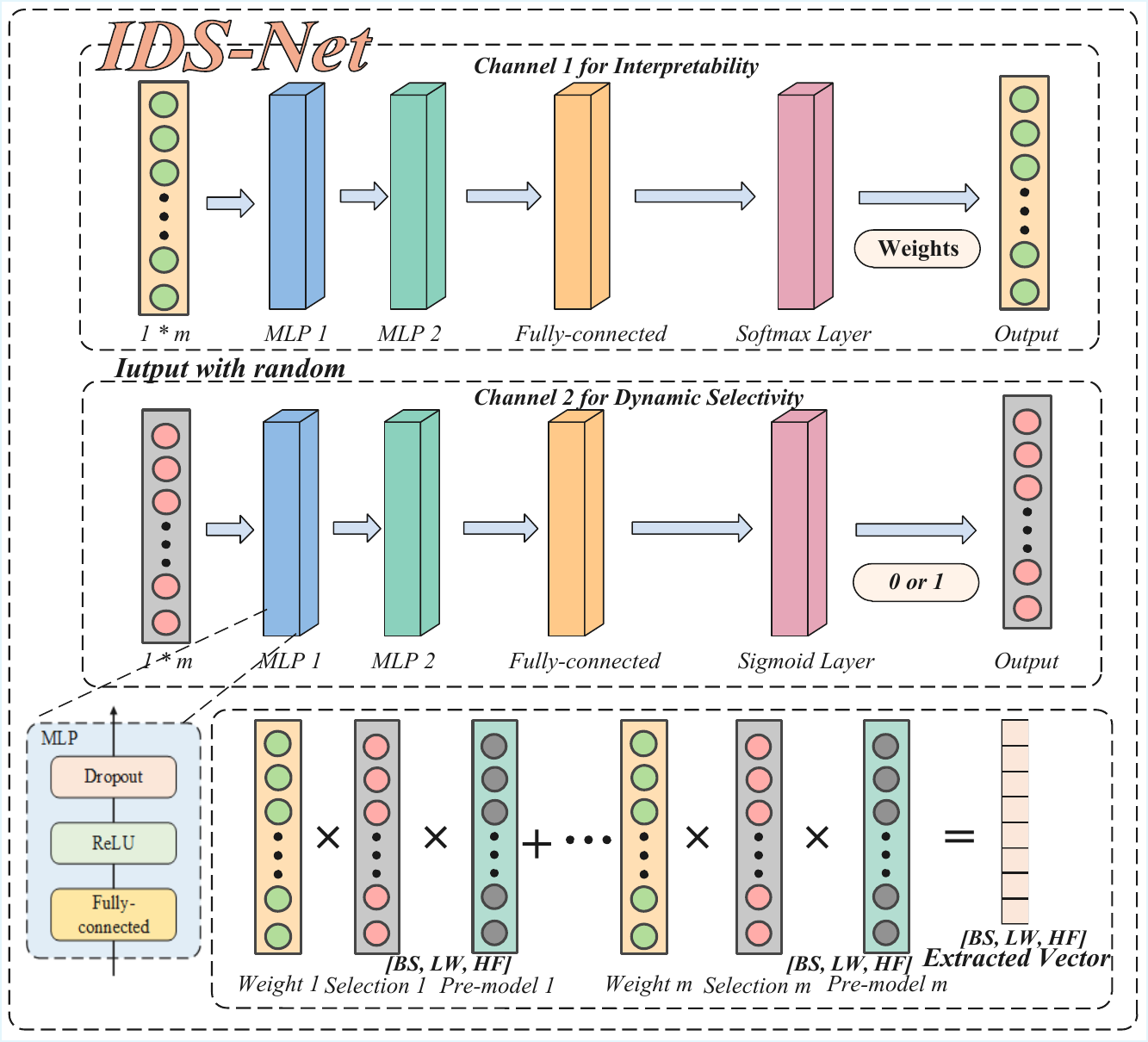}
    \vspace{-0.4\baselineskip}
    \caption{\textbf{Flow chart of the dual-channel parameter learning structure}}
\end{figure} 

\textit{\textbf{Channel 1 for interpretability}}: In Channel 1, our construct interpretability weight generation channel primarily consists of two MLP layers, a single fully connected layer, and a final Softmax layer. Specifically, we aim to train a set of weight values such that their sum equals 1, corresponding to the number of sub-models. A higher weight value signifies a more substantial contribution of the respective sub-model to the overall accuracy, thus facilitating effective model interpretation through these assigned weights. First, we generate a set of random numbers whose dimension matches the number of sub-models. These random numbers are then comprehensively learned and mapped into a high-dimensional space through a two-layer MLP and a fully connected layer. Finally, a softmax layer produces the specific weight outputs.

\vspace{0.25\baselineskip}

The MLP layer consists of a fully connected layer, a ReLU layer, and a dropout layer. The fully connected layer is used to extract data features and map the data from a low-dimensional space to a high-dimensional space for comprehensive data representation learning. Its formula is as follows:

\begin{equation}
    y_{i,j}^{l,fc} = \sum_{i=1}^{k} \omega_{i,j}^{l} \times y_{l-1,i} + b_{j}^{l},
\end{equation}
where $y_{i,j}^{l,fc}$ corresponds to the output of channel $j$ within the fully connected layer $l$. The variable $k$ specifies the total number of neurons present in layer $l-1$. Moreover, $\omega_{i,j}^{l}$ is defined as the weight parameter, and $b_{j}^{l}$ functions as a bias term, aiding the model's capacity to conform to the underlying data distribution.

\vspace{0.25\baselineskip}

The ReLU is a widely used activation function in artificial neural networks, primarily for its effectiveness in introducing non-linearity into the model. Mathematically, ReLU is defined as follows:

\begin{equation}
    y_{i,j}^{l,\text{Relu}} = \max \left(0, y_{i,j}^{l,fc}\right),
\end{equation}
where, $y_{i,j}^{l,\text{Relu}}$ represents the output of the ReLU activation function for channel $j$ at index $i$ in layer $l$. The term $y_{i,j}^{l,fc}$ denotes the input to the ReLU function, which is the output from the fully connected layer for channel $j$ at index $i$ in layer $l$. The $\max(0, \cdot)$ operation ensures that all negative values are clamped to zero, while positive values are passed through unchanged. Subsequently, Dropout regularization is utilized to mitigate model overfitting.

\vspace{0.25\baselineskip}

Finally, a set of random numbers generated by the two-layer MLP is passed through a fully connected layer to match the dimension of the sub-model count, and the final weight probabilities are then obtained via the Softmax layer. Mathematically, the Softmax layer is defined as follows:

\begin{equation}
    \sigma(\mathbf{z})_j = \frac{e^{z_j}}{\sum_{k=1}^{K} e^{z_k}} \quad \text{for } j=1, \dots, K,
\end{equation}
where the $e^{z_j}$ denotes the output of the fully connected layer, and $K$ denotes the output dimension, which is consistent with the number of sub-models. Higher weights indicate a greater contribution of the respective sub-model to the overall prediction accuracy. We leverage these varying weight values to enhance the interpretability of PV predictions.

\vspace{0.25\baselineskip}

\textit{\textbf{Channel 2 for dynamic selection}}: In Channel 2, we construct a deep learning network model similar to Channel 1, primarily aiming to train a set of 0 or 1 variables consistent with the number of sub-models. These variables are for dynamic, adaptive model selection. This approach avoids the subjective judgment often used in traditional ensemble learning models for selecting appropriate sub-models based on weight magnitudes, instead employing an end-to-end intelligent selection method to directly output the model selection results. Specifically, we utilize two MLP layers to deeply extract the data representation relationships of random numbers. A fully connected layer then adjusts the output dimension, and finally, a Sigmoid layer outputs the 0 or 1 variables. Its mathematical expression is as follows:

\begin{equation}
    \mathbf{y} = \sigma\left(\text{FC}\left(\text{MLP}_2\left(\text{MLP}_1(\mathbf{x})\right)\right)\right),
\end{equation}

\begin{equation}
\sigma(x) = \frac{1}{1+e^{-\lambda x}},
\end{equation}
where $\lambda$ denotes the penalty term, and $\sigma(x)$ denotes the Sigmoid function. We incorporate this additional penalty into the Sigmoid function, effectively transforming it into an approximate step function. In this situation, Channel 2 will exclusively produce binary outputs (0 or 1). Therefore, we use 1 to represent the selection of a sub-model and 0 to represent its discarding.

\subsubsection{The feature information interaction and fusion strategy}

In this section, we primarily introduce our proposed multi-feature information fusion learning strategy. Mainstream prediction methods typically combine features and PV power, feeding them together as input to the forecasting model. In contrast, this paper explores a novel feature fusion prediction paradigm, which separately maps PV power data and auxiliary feature data into a high-dimensional space, followed by cross-correlation embedding to obtain the information fusion values. \textbf{Figure 4} demonstrates this process.

\begin{figure}[!h]
    \centering
    \captionsetup{labelfont=bf}
    \includegraphics[width=1.0\linewidth]{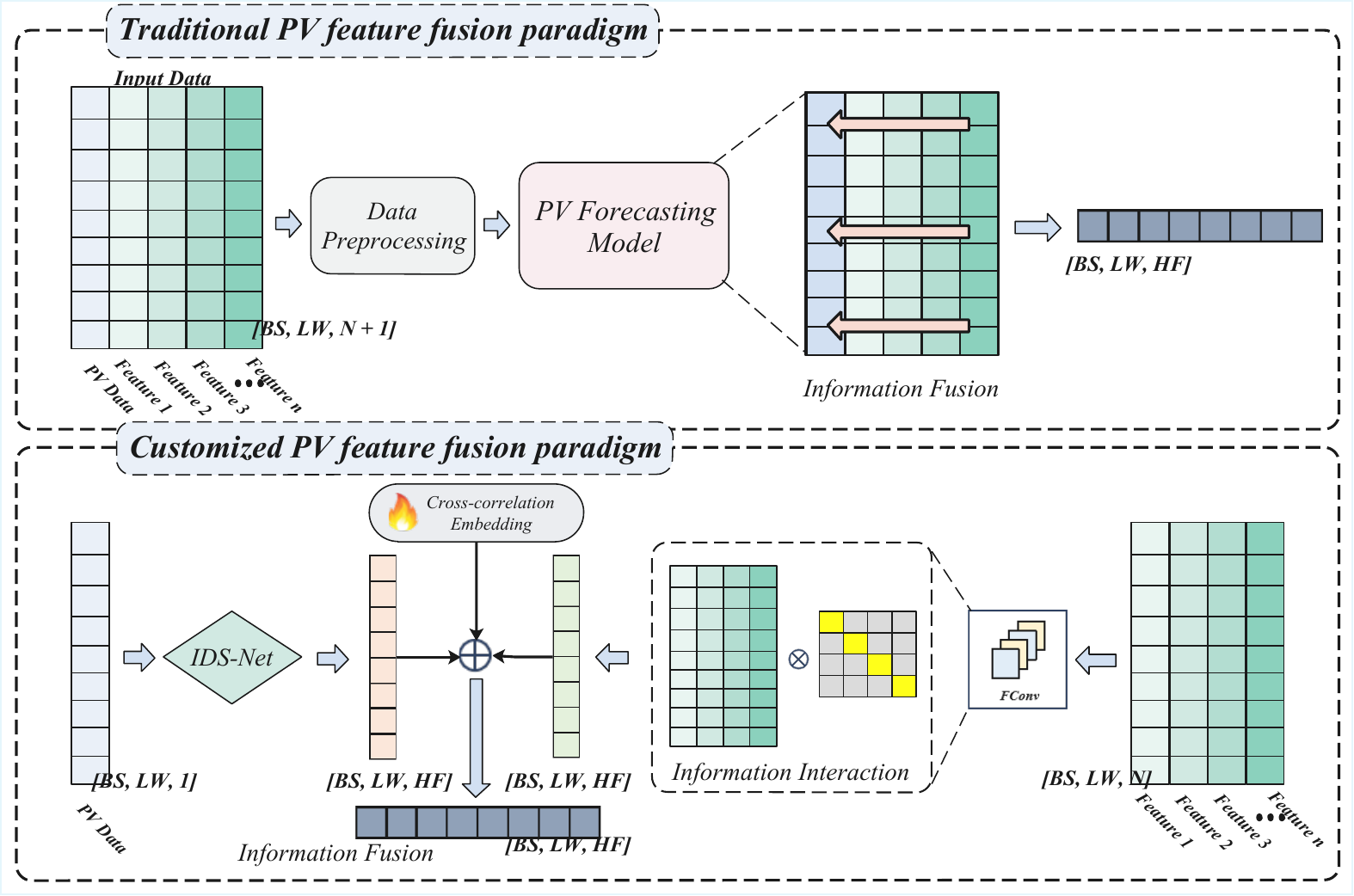}
    \vspace{-1.4\baselineskip}
    \caption{\textbf{Flow chart of feature interaction and fusion strategy}}
\end{figure} 

For additional PV-related feature variables, such as global horizontal irradiance and meteorological data, we first select the most relevant ones through the feature engineering method to reduce the computational burden caused by feature redundancy during model training. Then, we consider the additional PV features as $\mathbf{X} \in \mathbb{R}^{BS \times LW \times N}$, where $BS$ is the Batch Size, $LW$ is the Look-back Windows, and $N$ is the number of features. We aim to comprehensively capture the intricate inter-feature relationships using a feature convolutional network (FConv), thereby projecting the auxiliary feature variables into a high-dimensional representation. Its mathematical process can be defined as follows:

\begin{equation}
Input: [BS,\:LW,\:N]\overset{\text{FCnov}}{\rightarrow} After \ Interaction: [BS,\:LW,\:HF],
\end{equation}
where $HF$ denotes the High-dimension Feature. A similar data mapping approach is applied to PV power data. The mathematical expression for its high-dimensional space mapping process is as follows:

\begin{equation}
Input: [BS,\:LW,\:1]\overset{\text{IDS-Net}}{\rightarrow} Extrected: [BS,\:LW,\:HF].
\end{equation}

Finally, in the information fusion process, we set a learnable weight parameter $\lambda$ for information fusion. Mathematically, the process of cross-correlation embedding is as follows:

\begin{equation}
After \ Fusion = \lambda \times Extracted \ \ PV  + (1-\lambda) \times Extracted \ \ Features.
\end{equation}

In the traditional PV feature fusion paradigm, since PV data and features are concatenated into the matrix as input to the prediction model, it is difficult to ascertain the weight contribution of features to the prediction results. Furthermore, treating the contributions of different features to prediction accuracy as similar can easily lead to some features negatively impacting the prediction accuracy. Consequently, our devised tailored feature interaction and fusion strategy effectively achieves a balance between PV data and auxiliary feature extraction. Furthermore, feature interpretability is augmented via the integration of learnable parameters. The FConv network enables the model to learn relationships with PV data by adding different weights to the features.

\subsection{Customized loss function and transfer learning strategy}

In this section, we primarily introduce our customized deep learning loss function and a rapid transfer learning strategy. We enhance the model's ability to learn peaks and valleys by adding a penalty factor to the traditional MSELoss. Utilizing an end-to-end rapid transfer learning strategy, which combines pre-transfer judgment and fine-tuning, effectively reduces the model's computational complexity while ensuring prediction accuracy.

\subsubsection{Customized penalty loss function}

For PV power prediction, the time intervals during which peaks and valleys emerge tend to be relatively consistent. For instance, peak information generally manifests during daylight hours, whereas troughs typically occur at nighttime. Accurate forecasting of peak values is vital for grid operators to conduct efficient power dispatch. To this end, we attach a penalty factor to the traditional MSELoss. Specifically, we define daily PV power values deviating by two standard deviations from the mean as either peaks or troughs, and for these points, the calculated MSELoss incurs a three-fold loss penalty. The traditional MSELoss can be defined as follows:

\vspace{-0.4cm}
\begin{equation}
MSE = \frac{1}{M}\sum\limits_{t = 1}^M {\left( {{y_t} - {{\hat y}_t}} \right)^2} 
\end{equation}
where $M$ is the number of time points, ${y_t}$ is the true value, and ${{\hat y}_t}$ is the predicted value. The formula of the improved MSELoss can be defined as follows:

\begin{equation}
 IMSELoss = \frac{1}{M} \sum_{t=1}^{M} \text{loss}_t \ ,
\end{equation}

\begin{equation}
\text{loss}_i =
\begin{cases}
    3 \times \text{MSE}_t & \text{if } |y_t - \text{mean}(y)| > 2 \times \text{std}(y) \\
    \text{MSE}_t & \text{otherwise}
\end{cases}.
\end{equation}

\subsubsection{Rapid transfer learning strategy}

To enhance the prediction accuracy of newly constructed PV stations, we adopt the transfer learning strategy to rapidly migrate a prediction model pre-trained on similar large datasets to the new station for forecasting. Specifically, we select 100 time points from the small-sample dataset as test data to first directly evaluate the transfer performance of the pre-trained model on the few-shot dataset. Then, we established a prediction accuracy threshold (MAE = 0.1), aiming to ensure sufficient accuracy while minimizing the computational cost associated with additional training. If the pre-trained model's prediction accuracy on the small-sample test set meets this threshold, it will be directly adopted without fine-tuning. However, if the prediction accuracy falls below this threshold, we will use the remaining small-sample data to fine-tune the pre-trained model. Specifically, we employ a lower learning rate and fewer training epochs, which allows for both the preservation of original parameter information and rapid generalization to the new dataset. This method balances computational complexity and prediction accuracy, enabling the original model to quickly learn new data distribution characteristics. This strategy can be defined as:

\begin{equation}
    \text{Fine-Tuning Decision} = 
    \begin{cases} 
        \text{Fine-tune,} & \text{if MAE} > 0.2 \\
        \text{Do not fine-tune,} & \text{if MAE} \le 0.2
    \end{cases} \ \ \ .
\end{equation}

\section{Data Selection, Parameter Design, and Evaluation Index}

In this section, we primarily introduce four PV power station datasets from Hebei, China. Two of these stations have sufficient historical data, while the other two are newly constructed and lack enough training data. Additionally, this section also describes the parameters of the data processing module and the PV forecasting module, experimental evaluation metrics, and experimental environment information.

\subsection{Data selection}

In this study, to validate that our proposed transfer strategy and IDS-Net model can effectively transfer the PV prediction model from large-sample datasets to small-sample datasets, we select two newly established PV power station to verify their generalization ability. For model pre-training, we choose two PV power stations within the same PV farm that had the most similar data distribution and ample historical data. Next, we select the last 100 data points from the small-sample dataset as the test set to determine whether model fine-tuning is necessary. Finally, we apply the proposed strategy and forecasting model to the small-sample PV prediction problem to simulate its effectiveness in real-world PV small-sample prediction scenarios.

We selected suitable PV power data, recorded at 15-minute intervals, from the PV power station in Hebei Province, China. We selected two newly constructed power stations, station08 (from September 27 to September 30, 2018) and station02 (from April 10 to April 13, 2019), as objects for transfer learning. To improve the prediction accuracy for small sample sizes, we employed the MMD algorithm to select two pre-training stations with the most similar data distributions during the same period. These were station07 and station03, with historical data ranging from June 30 to September 30, 2018, and January 11 to April 13, 2019, respectively. The specific dataset distribution and the split into training and test sets are illustrated in  \textbf{Figure 5} and \textbf{Table 2}.

\begin{figure}[!h]
    \centering
    \captionsetup{labelfont=bf}
    \includegraphics[width=1.0\linewidth]{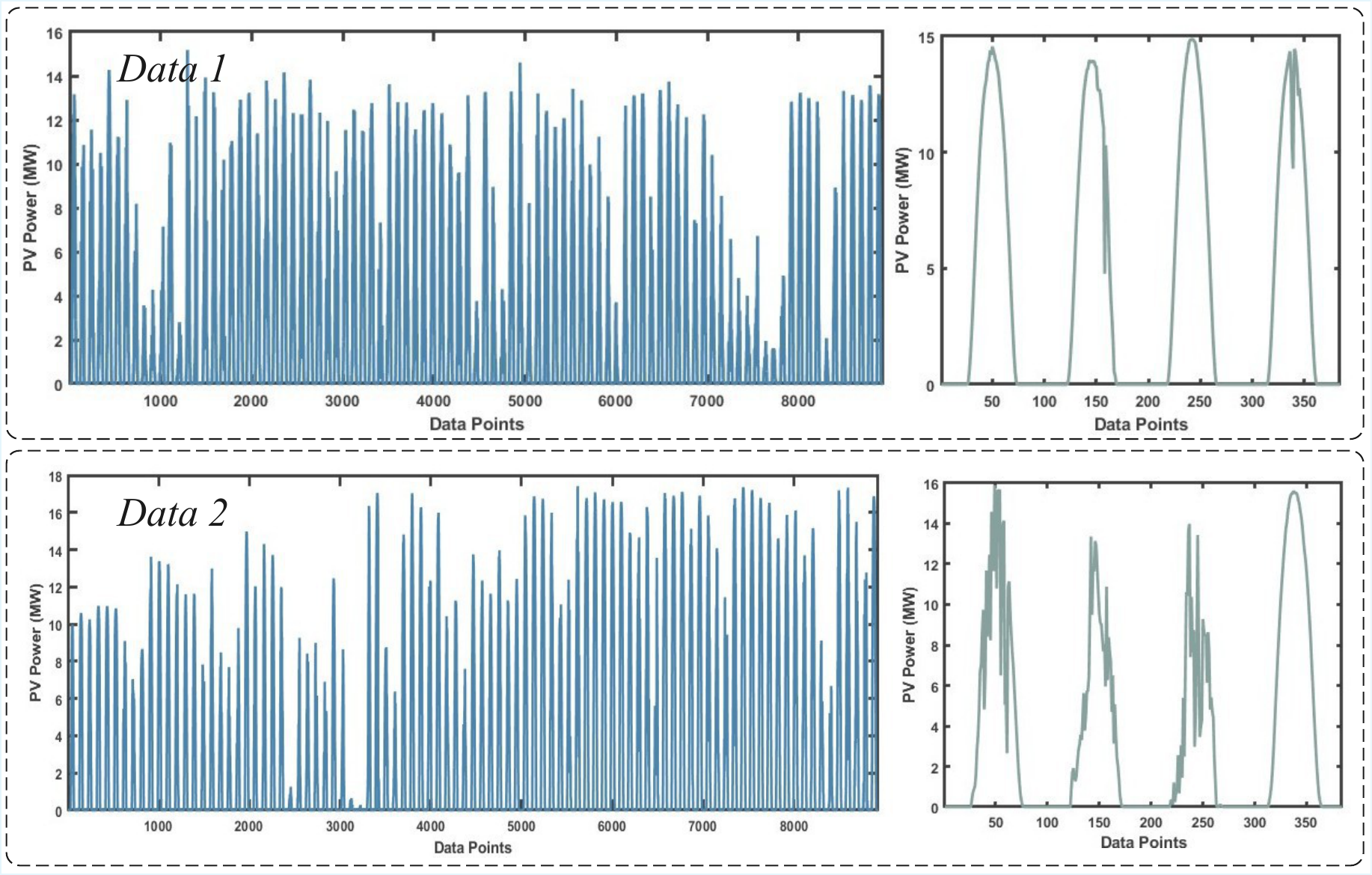}
    \vspace{-1.0\baselineskip}
    \caption{\textbf{The characteristics of the two different datasets}}
\end{figure}  

\begin{table}[h]
\small
\begin{spacing}{1.6}
\caption{\textbf{Basic characteristics of the different PV power datasets}}
\small
\setlength{\tabcolsep}{1.3mm}{
\begin{tabular}{ccccccccccc}
\hline
 &  & Data Points & Training Set & Validating Set& Testing Set & Mean  & Max    & Min  & Var     & Std   \\ \hline
\multirow{2}{*}{\textbf{Data 1}}  & Large Dataset & 8928  &8528 &300  &100  & 2.58    &15.17    &  0  &14.89     &  3.85 \\
  &Small Dataset  & 384  & 284  & \ding{55} & 100 & 4.28  & 14.86   & 0   &30.68     &5.53   \\\hline
\multirow{2}{*}{\textbf{Data 2}} & Large Dataset & 8928  &8528 &300  &100&  3.15& 17.42    &  0  & 24.06    & 4.91  \\
  &Small Dataset  & 384  & 284  & \ding{55} & 100 & 3.59  & 15.89   & 0   &24.39     & 4.93  \\\hline
\end{tabular}}
\end{spacing}
\end{table}

\subsection{Parameter design}
Our proposed PV forecasting framework is divided into two modules: the data preprocessing module and the PV forecasting module. In the data preprocessing module, we mainly adopt the MMD algorithm, the HI algorithm, and the ReliefF algorithm for data preprocessing. In the PV forecasting module, we use the dual-channel block and deep learning model pool for PV forecasting. In this study, the Grid Search is used to optimize the hyper-parameters of our framework. The hyper-parameters and inter-parameters are shown in \textbf{Table 3} and \textbf{Table 4}, and (*) demonstrates the parameters we adopt in this study.

\begin{table}[!h]
\small
\begin{spacing}{1.1}
\caption{\textcolor{DeBlue}{\textbf{Hyper-parameters of the proposed framework}}}
\small
\setlength{\tabcolsep}{3.4mm}{
\begin{tabular}{cccc}
\hline
& \textbf{Model}      & \textbf{Parameters} & \textbf{Value}   \\ \hline
\multirow{5}{*}{\textbf{Data Preprocessing Module}} & \multirow{1}{*}{\textbf{MMD}}    & $\sigma$   & Based on Grid Search     \\ \cline{2-4}
  &   \multirow{2}{*}{\textbf{HI}}     & Threshold       & 3             \\
   &                   & Window size    & 5            \\ \cline{2-4}
& \multirow{2}{*}{\textbf{ ReliefF}} & K (number of nearest neighbors)   &	70    \\
      &        & Distance metric       &	Euclidean  \\    \hline
\multirow{9}{*}{\textbf{Forecasting Module}}      
& \multirow{7}{*}{\textbf{IDS-Net}}  & Look-back window    &[32, 64, 96*, 128]         \\
 &               & Time step feature   & Based on ReliefF \\
 &                    & Target length       & 1          \\
&        & Learning rate       & 0.0001       \\
 &                 & Optimizer   & AdamW       \\
      &         & Max epoch      & [100, 200, 300*, 500]             \\
     &              & Batch size          &  [32, 64, 128*, 256]               \\ \cline{2-4}
& \multirow{2}{*}{\textbf{Fine-tuning}} & Max epoch & 50      \\
     &              & Batch size          &  16         \\ \hline
\end{tabular}}
\end{spacing}
\end{table}

\begin{table}[!h]
\small
\begin{spacing}{1.1}
\caption{\textbf{Inter-parameters of the deep learning model pool}}
\small
\setlength{\tabcolsep}{17mm}{
\begin{tabular}{ccc}
\hline
 \textbf{Model}      & \textbf{Parameters} & \textbf{Value}   \\ \hline
 \multirow{2}{*}{\textbf{LR}} &    Layer number       &      2    \\  
 & Hidden size   &  64            \\ \hline
  \multirow{3}{*}{\textbf{MLP}} &    Hidden size   &  64   \\    
  &  Activation function    & ReLU   \\ 
    &  Regularization    & Dropout   \\ \hline
   \multirow{4}{*}{\textbf{CNN}}&  Layer number    &   2           \\
 &    Kernel size       &   3    \\
  &    Padding   &    1         \\
 &    Pooling stride     & 2            \\  \hline
\multirow{2}{*}{\textbf{GRU}} &    Layer number       &      2    \\  
 &  Hidden size	        & 64   \\ \hline
\multirow{2}{*}{\textbf{RNN}} &    Layer number       &      2    \\  
 &  Hidden size	        & 64   \\ \hline
 \multirow{2}{*}{\textbf{LSTM}} &    Layer number       &      2    \\  
 &  Hidden size	        & 64   \\ \hline
 \multirow{2}{*}{\textbf{BiLSTM}} &    Layer number       &      2    \\  
 &  Hidden size	        & 64   \\ \hline
 \multirow{3}{*}{\textbf{BiLSTM-Attention}} &    Layer number  &   2    \\  
 &  Hidden size	        & 64   \\ 
  &  Number heads	        & 16   \\ \hline
\end{tabular}}
\end{spacing}
\end{table}

\subsection{Evaluation indexes}
To comprehensively evaluate the performance of our proposed model in PV power forecasting, we introduce several evaluation metrics to measure the disparity between the true values and the predicted values. The evaluation metrics are Mean Squared Error (MSE), Mean Absolute Error (MAE), Root Mean Squared Error (RMSE), R-Square (${R}^{2}$), which can be defined as follows:

\begin{equation}
    \text{MSE} =  \frac{1}{n} \sum_{i=1}^{n} (\hat{f}_{mi} - f_{mi})^2
\end{equation}

\begin{equation}
    \text{MAE} = \frac{1}{n} \sum_{i=1}^{n} |\hat{f}_{mi} - f_{mi}|
\end{equation}

\begin{equation}
    \text{RMSE} = \sqrt{\frac{1}{n} \sum_{i=1}^{n} (\hat{f}_{mi} - f_{mi})^2}
\end{equation}

\begin{equation}
    R^2 = 1 - \frac{\sum_{i=1}^{n} (\hat{f}_{mi} - f_{mi})^2}{\sum_{i=1}^{n} (f_{mi} - \bar{f}_{mi})^2}
\end{equation}

\subsection{Operation environment}
The basic operating environment of the proposed IDS-Net framework is Intel i7-9700 CPU @ 3.00 GHz, 16 GB RAM, and 4 NVIDIA GeForce RTX 3090. Our work utilizes MATLAB 2022a for the data preprocessing module. Our proposed model is also constructed using PyTorch with the assistance of PyCharm IDE.

\section{Numerical Verification}

In this section, we comprehensively validate the effectiveness of the transfer strategy in our proposed PV forecasting framework on small-sample datasets through four experiments, effectively addressing the limited data problem for newly constructed PV power stations. First, we validate the superior performance of our proposed IDS-Net in PV power forecasting by comparing it with traditional deep learning models and state-of-the-art time series forecasting models. Second, we confirm the necessity of each component of our proposed IDS-Net framework through ablation experiments. Furthermore, we also verified the effectiveness of our proposed rapid transfer learning strategy and conducted a sensitivity analysis to confirm the potential engineering applicability of our proposed PV forecasting framework.

\subsection{Remarkable performance of our proposed framework}

In this section, we primarily evaluate the performance of various time series forecasting models on both large-scale dataset prediction and small-sample transfer prediction. These include representatives of traditional deep learning, such as MLP, CNN, GRU, and LSTM, as well as state-of-the-art time series forecasting models, such as Transformer, DLinear \cite{zeng2023transformers}, and PatchTST \cite{nie2022time}. To facilitate a fair and rigorous assessment, each comparison model was trained on an identical feature set, with its hyperparameters finely tuned to achieve optimal results. The comparison results of different models are presented in \textbf{Figure 6} and \textbf{Figure 7}, with specific results provided in \textbf{Table 5}.

\begin{figure}[!h]
    \centering
    \captionsetup{labelfont=bf}
    \includegraphics[width=1.0\linewidth]{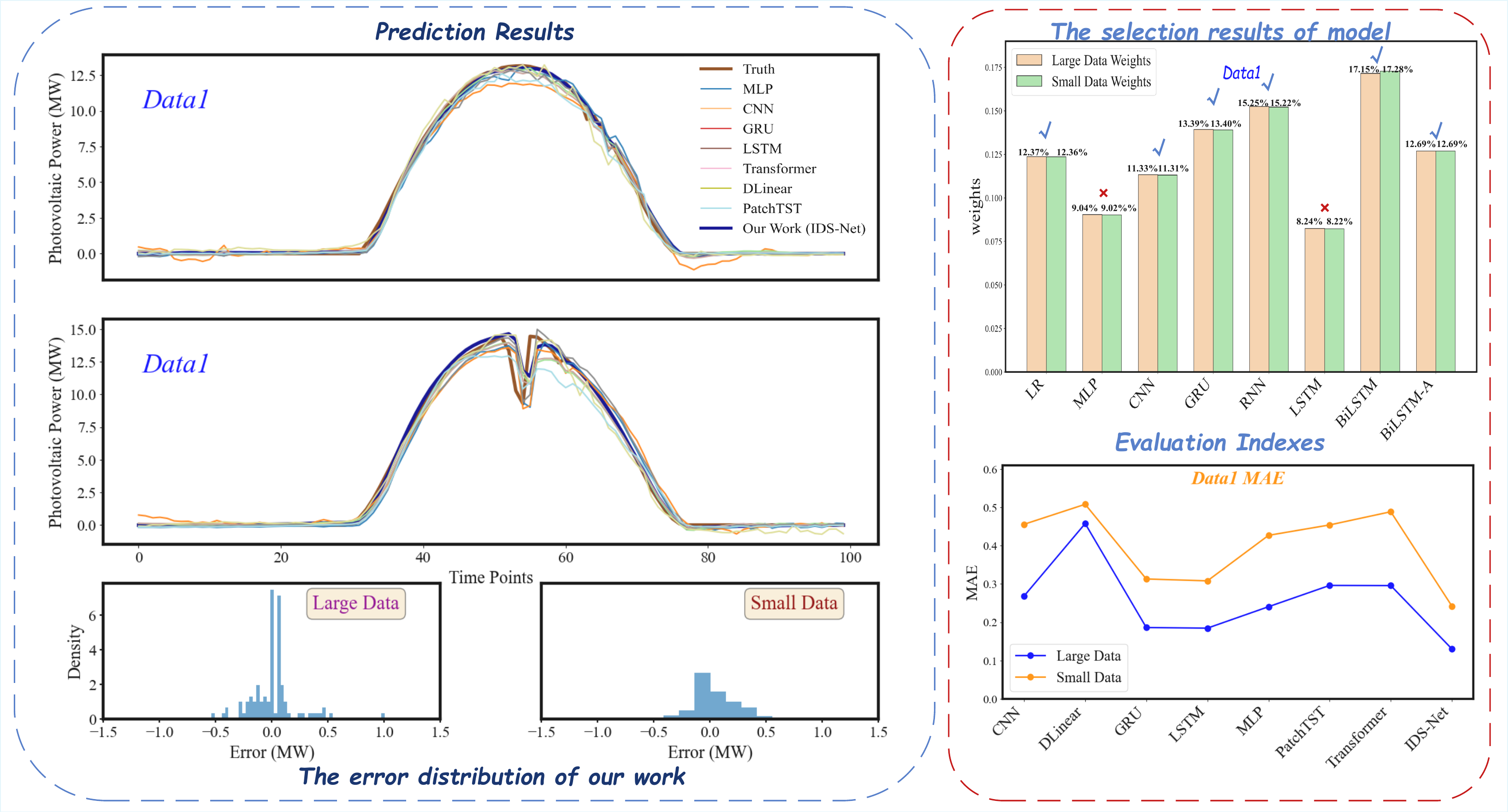}
    \vspace{-0.6\baselineskip}
    \caption{\textbf{Prediction results of different models on Data 1}}
\end{figure}

\begin{figure}[!h]
    \centering
    \captionsetup{labelfont=bf}
    \includegraphics[width=1.0\linewidth]{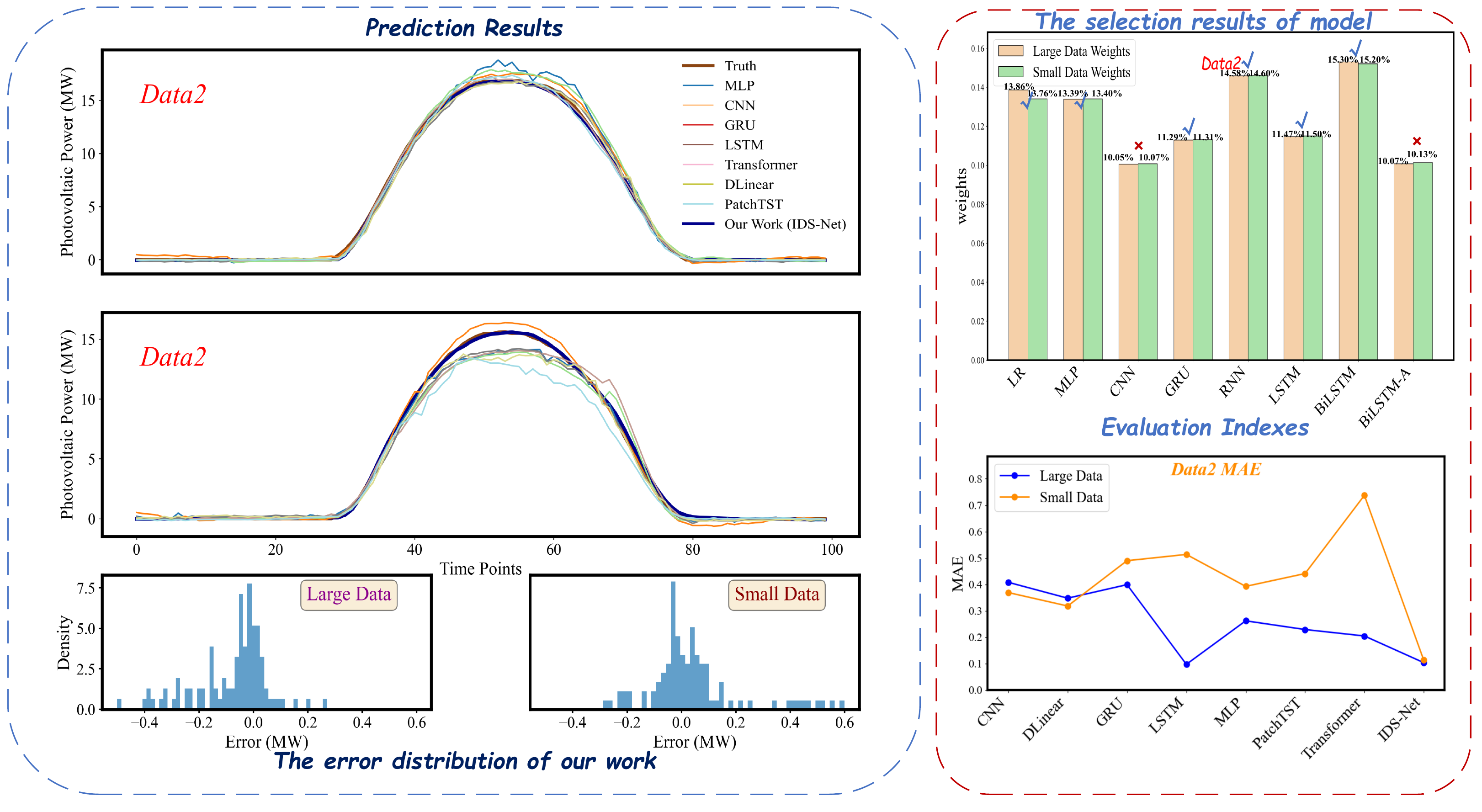}
    \vspace{-0.6\baselineskip}
    \caption{\textbf{Prediction results of different models on Data 2}}
\end{figure}

In Data 1, our proposed IDS-Net model demonstrates superior performance across both large and small datasets, serving as a compelling comparison against traditional deep learning models and state-of-the-art forecasting models. For the large dataset, our work significantly reduces prediction errors. Compared to MLP, our model lowers MSE by an impressive 73.66\%, dropping from 0.1299 to 0.0357, indicating a substantial improvement in overall accuracy by minimizing the squared differences between predicted and actual values. Against the Transformer model, our work achieves a remarkable 60.68\% reduction in MAE, moving from 0.2958 to 0.1164, signifying a more accurate average magnitude of errors. Furthermore, our $R^2$ value increases by 0.75\% (from 0.9917 to 0.9986) when compared to Transformer, indicating that our model explains a larger proportion of the variance in the target variable, demonstrating superior generalization capability. Our proposed model also achieves state-of-the-art (SOTA) performance when compared to several advanced Transformer-based models and linear models. Notably, against DLinear, a recent linear model, our work achieves an 89.44\% improvement in MSE, reducing it from 0.3369 to 0.0357, further confirming our model's superiority. Moreover, when fine-tuned on the small dataset, our proposed IDS-Net continues to exhibit outstanding performance, showcasing its effectiveness in transfer learning to low-data regimes. Compared to MLP, our model reduces MSE by 37.19\% (from 0.5792 to 0.3648). Similarly, against Transformer, our work achieves a substantial 50.70\% reduction in MAE (from 0.4887 to 0.2422) and an increase in $R^2$ by 0.51\% (from 0.9770 to 0.9872). These consistent improvements across different evaluation metrics and dataset sizes comprehensively demonstrate the effectiveness and superiority of our proposed IDS-Net model.

\vspace{0.25\baselineskip}

In Data 2, our proposed IDS-Net model continues to demonstrate superior performance across both large and small datasets, further validating its robustness and generalization capabilities. For the large dataset in Data 2, our work significantly reduces prediction errors across all metrics. Compared to MLP, our model lowers MSE by an impressive 84.19\%, dropping from 0.1474 to 0.0234, indicating a substantial improvement in accuracy. Against the Transformer model, our work achieves a remarkable 49.34\% reduction in MAE, moving from 0.2045 to 0.1034. Furthermore, our $R^2$ value shows a slight increase from 0.9981 (Transformer) to 0.9995, indicating that our model explains an even larger proportion of the variance in this new dataset. Notably, when compared to DLinear, our work achieves an 87.59\% improvement in MSE, reducing it from 0.1884 to 0.0234, further solidifying our model's superiority over linear approaches. Moreover, when fine-tuned on the small dataset of Data 2, our proposed IDS-Net maintains its outstanding performance. Compared to MLP, our model reduces MSE by 74.32\% (from 0.1210 to 0.0315). Similarly, against Transformer, our work achieves a substantial 84.81\% reduction in MAE (from 0.7371 to 0.1129) and an increase in $R^2$ from 0.9599 to 0.9954. These consistent and significant improvements across Data 2, replicating the trends observed in Data 1, comprehensively demonstrate the effectiveness, robustness, and superior generalization capability of our proposed IDS-Net model in diverse data environments, particularly in both large-scale pre-training and small-sample fine-tuning scenarios.

\vspace{0.25\baselineskip}

\textbf{Remark 1.} Experiment 1 validated the effectiveness of our proposed IDS-Net model. It consistently outperformed both traditional deep learning models and the latest state-of-the-art forecasting models across all experimental metrics and on both large pre-training datasets and small fine-tuning datasets. Furthermore, by leveraging two distinct datasets (Data 1 and Data 2), we simultaneously verified the model's robust generalization capability and its superior performance in diverse datasets.

\newpage
\newgeometry{landscape, left=1.5cm, right=3cm,top=0.5cm,bottom=0.5cm}
\begin{landscape}
\pdfpagewidth=210mm 
\pdfpageheight=297mm 
\begin{table}[]
\begin{spacing}{1.2}
\caption{\textbf{Prediction results of two datasets in Experiment 1}}
\setlength{\tabcolsep}{5.4mm}
\begin{tabular}{llllllllll}
\hline
&  &\multicolumn{4}{l}{\textbf{Large Dataset}} &\multicolumn{4}{l}{\textbf{Small Dataset}} \\ \hline
 &    & \textbf{MSE}     & \textbf{MAE}  & \textbf{RMSE}     & \textbf{$R^2$} & \textbf{MSE}     & \textbf{MAE}  & \textbf{RMSE}     & \textbf{$R^2$} \\ \hline
\multirow{8}{*}{\textbf{Data 1}} & MLP    & 0.1299   &0.2406  & 0.3605  & 0.9948  & 0.5792 & 0.4270 &  0.7611&0.9797 \\
& CNN    &0.1753&0.2684&0.4187& 0.9930& 0.6448  &0.4559  &0.8030 & 0.9774 \\
& GRU    & 0.0666 & 0.1862 & 0.2581 & 0.9973 & 0.4219  &0.3130  &0.6496 & 0.9852 \\
& LSTM  &  0.0709  & 0.1846   & 0.2663 & 0.9972  &  0.3959  &  0.3081 &  0.6292   & 0.9861    \\
& Transformer    &0.2092    & 0.2958  &  0.4574    & 0.9917  & 0.6540   & 0.4887   & 0.8087 & 0.9770  \\
& DLinear     & 0.3369  & 0.4578  &  0.5805   &0.9866 & 0.6900  &  0.5081   & 0.8306 &  0.9758\\
& PatchTST     &  0.1873  &0.2962   & 0.4328   &  0.9925 & 0.5983   &  0.4542  & 0.7735 & 0.9790  \\
& \textbf{Our Work (IDS-Net)}&  \textbf{0.0357}  & \textbf{0.1164} & \textbf{0.1888}  & \textbf{0.9986}  & \textbf{0.3648}  &\textbf{0.2422}  & \textbf{0.6040}&\textbf{0.9872 } \\ \hline
\multirow{8}{*}{\textbf{Data 2}} & MLP    &  0.1474  & 0.2776 & 0.3839  &  0.9967 &  0.1210 & 0.2693 &0.3478 & 0.9967 \\
& CNN    & 0.1207   & 0.2472 &  0.3474  &  0.9973 & 0.2709  &0.3823  & 0.5205 &  0.9926\\
& GRU    & 0.1871   &0.3320  & 0.4325  & 0.9958  &  0.3519 &  0.3082&0.5932 & 0.9904 \\
& LSTM    &  0.1548  & 0.3003 &0.3935   &  0.9965 & 0.3141  & 0.3530 & 0.5605 &0.9914  \\
& Transformer    &0.0847  & 0.2045 & 0.2910  &  0.9981& 1.4647  &0.7371 &1.2103 &0.9599  \\
& DLinear    & 0.1884  &  0.3476 & 0.4340  &0.9957   & 0.1788  &0.3179 &  0.4229& 0.9951 \\
& PatchTST    & 0.1656   & 0.3002 & 0.4069  &0.9963   &  0.2270 &0.3428  & 0.4765& 0.9938 \\
& \textbf{Our Work (IDS-Net)}    & \textbf{0.0234}   & \textbf{0.1034} & \textbf{0.1530}  &\textbf{0.9995}   &\textbf{0.0315}   &\textbf{0.1129}  &\textbf{0.1774} & \textbf{0.9991} \\ \hline
\end{tabular}
\end{spacing}
\end{table}
\end{landscape}
\restoregeometry

\subsection{The ablation experiment of our proposed framework}

In this section, we conduct ablation studies on the proposed PV forecasting framework to validate the necessity of each introduced block. Specifically, we investigate four key components: (1) we examine the impact of feature engineering by removing it and fusing all raw dataset features; (2) we assess our customized feature fusion paradigm by replacing it with a traditional approach; (3) for the ensemble learning module, we evaluate our proposed dual-channel weighted ensemble strategy against a simple average weighting of sub-models; and (4) we analyze the contribution of our customized loss function strategy by training the model solely with a traditional MSELoss. The experimental results are shown in \textbf{Table 6}.

\vspace{0.25\baselineskip}

In Data 1, we conducted ablation studies to validate the necessity of each introduced block in our proposed IDS-Net model. The results consistently demonstrate that each component contributes positively to the overall performance, both on the large pre-training dataset and the small fine-tuning dataset. For the large dataset, removing any of our proposed components led to a performance degradation. Specifically, the model "w/o Feature Engineering" saw its MSE increase from our model's 0.0357 to 0.0564, and its MAE from 0.1164 to 0.1534, highlighting the crucial role of our feature engineering module. Similarly, without our customized feature fusion, the MSE increased to 0.0400 and the MAE to 0.1349. Disabling our two-stage weighted prediction resulted in an MSE of 0.0649 and an MAE of 0.1425. Lastly, training without our customized loss function led to an MSE of 0.0554 and an MAE of 0.1632. In all these ablation scenarios, the performance metrics (MSE, MAE, RMSE) worsened, and $R^2$ decreased compared to our full IDS-Net model, confirming the necessity of each module. Similar trends were observed for the small dataset. For instance, "w/o Feature Engineering" increased MSE to 0.4458 and MAE to 0.3150 compared to our model's 0.3648 and 0.2422, respectively. "w/o Feature Fusion" also resulted in higher errors (MSE: 0.3802, MAE: 0.2502). The absence of our "Model Weighted Prediction" led to an MSE of 0.4326 and an MAE of 0.2776. Finally, "w/o Customized Loss Function" yielded an MSE of 0.4225 and an MAE of 0.2927. The consistent degradation in performance across all ablation scenarios on both large and small datasets robustly confirms the indispensable contribution of each proposed block to the superior performance of our IDS-Net model.

\vspace{0.25\baselineskip}

The ablation studies in Data 2 further corroborate the integral role of each component within our proposed IDS-Net model. For the large dataset, the absence of "Feature Engineering" resulted in an increased MSE from our model's 0.0234 to 0.0266 and MAE from 0.1034 to 0.1233, underscoring the benefits of our engineered features. Similarly, without "Feature Fusion," the MSE rose to 0.0238 and MAE to 0.1143, validating the efficacy of our fusion approach. Furthermore, training without the "Customized Loss Function" resulted in an MSE of 0.0422 and MAE of 0.1672, confirming its optimization advantage. Consistent with these findings, the small dataset in Data 2 also exhibited performance drops under ablation. For instance, the "w/o Feature Fusion" variant saw its MSE rise to 0.1387 and MAE to 0.2542. The absence of "Model Weighted Prediction" led to an MSE of 0.2465 and an MAE of 0.3234. Finally, "w/o Customized Loss Function" yielded an MSE of 0.3791 and MAE of 0.3966. The uniform and substantial performance degradation was observed across all ablation scenarios on both large and small datasets.

\vspace{0.25\baselineskip}
\textbf{Remark 2.} Experiment 2 validates the necessity of the designed blocks within our proposed IDS-Net framework, which demonstrates that our proposed blocks can effectively enhance the accuracy and robustness of our model.

\newpage
\newgeometry{landscape, left=1.5cm, right=3cm,top=0.5cm,bottom=0.5cm}
\begin{landscape}
\pdfpagewidth=210mm 
\pdfpageheight=297mm 
\begin{table}[]
\begin{spacing}{1.4}
\caption{\textbf{Prediction results of two datasets in Experiment 2}}
\setlength{\tabcolsep}{4.4mm}
\begin{tabular}{llllllllll}
\hline
&  &\multicolumn{4}{l}{\textbf{Large Dataset}} &\multicolumn{4}{l}{\textbf{Small Dataset}} \\ \hline
 &    & \textbf{MSE}     & \textbf{MAE}  & \textbf{RMSE}     & \textbf{$R^2$} & \textbf{MSE}     & \textbf{MAE}  & \textbf{RMSE}     & \textbf{$R^2$} \\ \hline
\multirow{5}{*}{\textbf{Data 1}}   & w/o Feature Engineering    &   0.0564 &0.1534  & 0.2375  & 0.9978   &0.4458   &0.3150  & 0.6677 &0.9843  \\
& w/o Feature Fusion    &0.0400    & 0.1349 &  0.2001 &0.9984   & 0.3802  & 0.2502 &0.6166 &0.9866  \\
& w/o Model Weighted Prediction    & 0.0649   & 0.1425 &0.2548   & 0.9974  & 0.4326  & 0.2776 &0.6577 &0.9848  \\
& w/o Customized Loss Function    & 0.0554   & 0.1632 & 0.2354  & 0.9978  & 0.4225  &  0.2927&0.6500  &  0.9852\\
& \textbf{Our Work (IDS-Net)}&  \textbf{0.0357}  & \textbf{0.1164} & \textbf{0.1888}  & \textbf{0.9986}  & \textbf{0.3648}  &\textbf{0.2422}  & \textbf{0.6040}&\textbf{0.9872 } \\ \hline
\multirow{5}{*}{\textbf{Data 2}}   & w/o Feature Engineering    & 0.0266   & 0.1233 &   0.1632 & 0.9994  & 0.1923  & 0.2895 & 0.4386&0.9947  \\
& w/o Feature Fusion    &0.0238    & 0.1143 & 0.1542  &0.9995   &  0.1387 &0.2542  & 0.3725 &0.9962  \\
& w/o Model Weighted Prediction    &  0.0424  & 0.1474 &  0.2059  &  0.9990  &0.2465   & 0.3234 &0.4965 & 0.9933 \\
& w/o Customized Loss Function    & 0.0422   &  0.1672&  0.2053 & 0.9990  &  0.3791 & 0.3966 & 0.6157&  0.9896\\
& \textbf{Our Work (IDS-Net)}    & \textbf{0.0234}   & \textbf{0.1034} & \textbf{0.1530}  &\textbf{0.9995}   &\textbf{0.0315}   &\textbf{0.1129}  &\textbf{0.1774} & \textbf{0.9991} \\ \hline
\end{tabular}
\end{spacing}
\end{table}
\end{landscape}
\restoregeometry

\subsection{The effectiveness of our proposed transfer strategy}

In this section, the significance of our transfer strategy is corroborated by this experiment, which establishes the greater effectiveness of adapting a pre-trained model from the source domain over training a new model directly with the few-shot samples. In this experiment, the last 100 data points of the few-shot dataset are used as the test set to evaluate the effectiveness of the transfer strategy. In this experimental setup, 'direct prediction' refers to training a new forecasting model solely on the few-shot dataset, while 'transfer learning' involves directly transferring the pre-trained model to the target domain without fine-tuning. The experimental results are shown in \textbf{Figure 8} and \textbf{Table 7}.

\begin{figure}[!h]
    \centering
    \captionsetup{labelfont=bf}
    \includegraphics[width=1.0\linewidth]{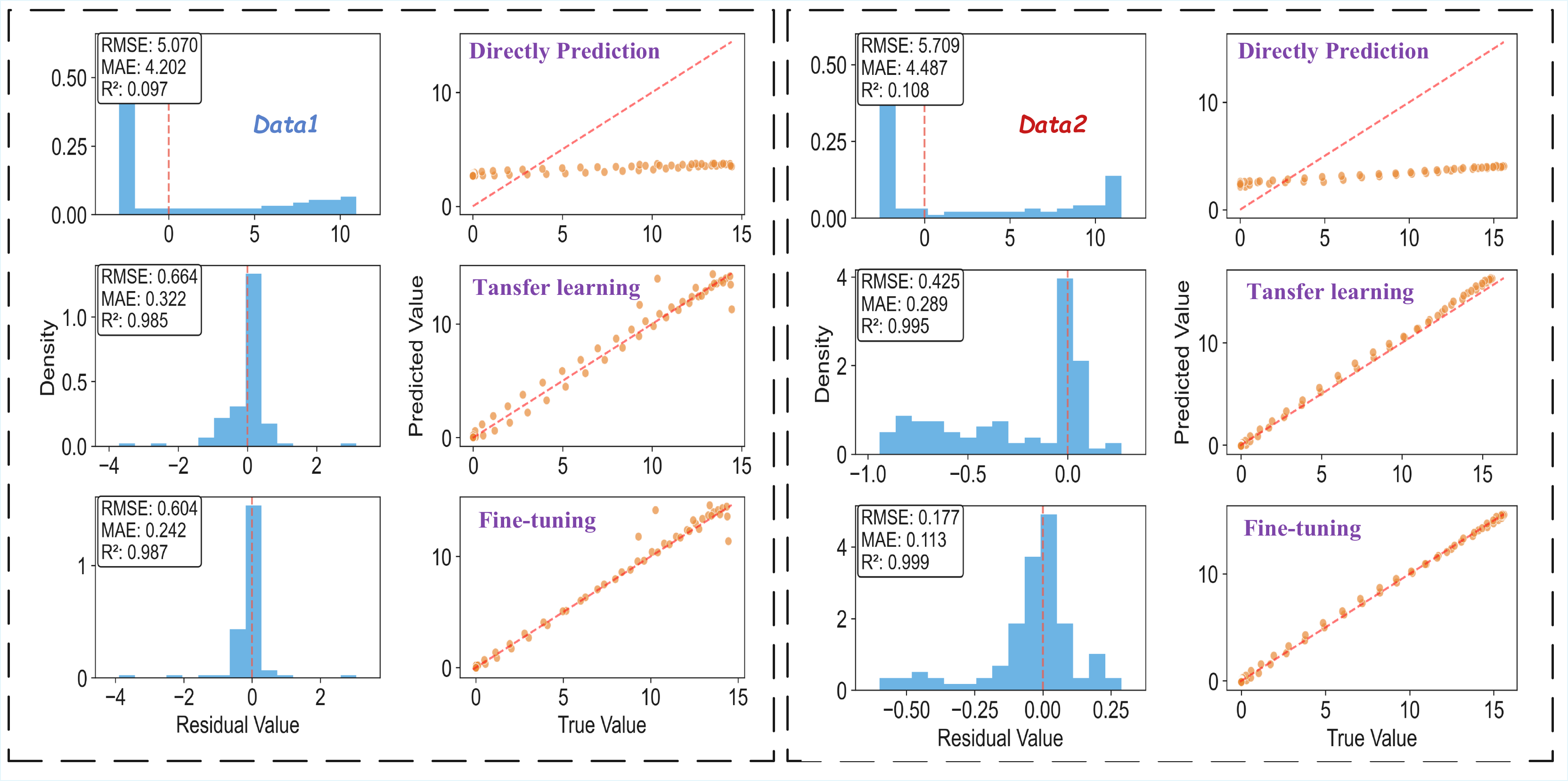}
    \vspace{-0.6\baselineskip}
    \caption{\textbf{Prediction performance of transfer learning}}
\end{figure}

\begin{table}[!h]
\begin{spacing}{1.4}
\caption{\textbf{Prediction results of two datasets in Experiment 3}}
\setlength{\tabcolsep}{5.8mm}
\begin{tabular}{llllll}
\hline
&  &\multicolumn{4}{l}{\textbf{Small Dataset}}\\ \hline
 &    & \textbf{MSE}     & \textbf{MAE}  & \textbf{RMSE}     & \textbf{$R^2$}  \\ \hline
\multirow{3}{*}{\textbf{Data 1}}  & Directly Prediction    &25.7098   & 4.2021    &5.0705 &0.0969 \\
& Transfer Learning   & 0.4412  &   0.3221 &0.6643 &0.9845 \\
& \textbf{Our Work (Fine-tuning)}  &  \textbf{0.3648}  &\textbf{0.2422}  & \textbf{0.6040}&\textbf{0.9872 }  \\  \hline
\multirow{3}{*}{\textbf{Data 2}}  & Directly Prediction    &32.5947   &  4.4871   &5.7092 &0.1083 \\
& Transfer Learning   & 0.1804  &  0.2886   &0.4247 &0.9951 \\
& \textbf{Our Work (Fine-tuning)}   &\textbf{0.0315}   &\textbf{0.1129}  &\textbf{0.1774} & \textbf{0.9991}\\  \hline
\end{tabular}
\end{spacing}
\end{table}

In Data 1, our proposed IDS-Net model, specifically optimized with fine-tuning, demonstrates superior performance in PV forecasting tasks on the small dataset, significantly outperforming both direct prediction and basic transfer learning strategies. For instance, in terms of MSE, our work achieves a remarkably low MSE of 0.3648, substantially outperforming the Directly Prediction approach (MSE 25.7098), representing a 98.58\% reduction in MSE. This highlights the critical role of pre-training and subsequent fine-tuning on limited data. To further validate the effectiveness of our fine-tuning strategy, we compare it against a basic Transfer Learning approach where no fine-tuning is performed. Our work achieves the lowest errors across key metrics, with an MSE of 0.3648, MAE of 0.2422, and RMSE of 0.6040. This substantially outperforms the Transfer Learning model (MSE 0.4412, MAE 0.3221, RMSE 0.6643), corresponding to a 17.31\% lower MSE and a 24.80\% lower MAE. Furthermore, our fine-tuned model achieves a higher $R^2$ of 0.9872, compared to Transfer Learning's 0.9845 and Directly Prediction's 0.0969. This consistent advantage of our fine-tuned IDS-Net model across all evaluation metrics underscores its superior predictive capability and the efficacy of our comprehensive transfer learning strategy for few-shot forecasting.

\vspace{0.25\baselineskip}

Similarly, in Data 2 of Experiment 3, our proposed IDS-Net model with fine-tuning consistently showcases its superior performance in PV forecasting tasks on the small dataset, further validating its robustness across different datasets. Our fine-tuned model achieves an exceptionally low MSE of 0.0315, drastically outperforming the Directly Prediction approach (MSE 32.5947), representing a 99.90\% reduction in MSE. This profound improvement reaffirms the crucial benefits of our integrated pre-training and fine-tuning methodology. When compared against the basic Transfer Learning strategy, our work maintains its lead across all key metrics. Our model achieves an MSE of 0.0315, MAE of 0.1129, and RMSE of 0.1774. This significantly outperforms the Transfer Learning model (MSE 0.1804, MAE 0.2886, RMSE 0.4247), corresponding to an 82.54\% lower MSE and a 60.95\% lower MAE. Moreover, our fine-tuned model yields a higher $R^2$ of 0.9991, compared to Transfer Learning's 0.9951 and Directly Prediction's 0.1083. This consistent and even more pronounced advantage of our full IDS-Net model with fine-tuning across all evaluation metrics in Data 2 further reinforces its superior predictive capability and the robust efficacy of its integrated transfer learning approach for few-shot forecasting challenges.

\vspace{0.25\baselineskip}

\textbf{Remark 3.} This experiment validates the effectiveness of our proposed transfer strategy. It highlights that leveraging knowledge from the pre-trained model and fine-tuning it on the target domain significantly improves the forecasting accuracy.

\subsection{Sensitive analysis}

In this section, we primarily evaluate our proposed model's practical engineering applicability. This is because, in real-world engineering, analyzing a model's sensitivity to hyperparameters is an effective method for assessing its robustness. Since deep learning prediction models heavily rely on hyperparameter tuning, achieving good prediction results even under sub-optimal parameters is also crucial. To effectively evaluate our model's sensitivity, we designed a single-factor sensitivity test. This method allows us to assess the impact of each hyperparameter on model performance and thus measure the sensitivity of our proposed framework by varying only one hyperparameter at a time while keeping others constant.

\vspace{0.25\baselineskip}

The hyperparameters primarily investigated in our work are Max Epoch (ME), Look-back Windows (LW), and Batch Size (BS). In this experiment, we set $\textbf{\textit{ME}}\in [100,\text{ }200,\text{ }{{300}^{*}},\text{ }500]$, $\textbf{\textit{LW}}\in [32,\text{ }{{64}},\text{ }{96}^{*},\text{ }128]$, and $\textbf{\textit{BS}}\in [32,\text{ }{{64},\text{ }128^{*}},\text{ }256]$, where the asterisk (*) denotes the optimal parameters in our framework. Specifically, the $\textbf{\textit{ME}}\Phi (\textbf{\textit{LW}}, \textbf{\textit{BS}})$ indicates changing the Max Epoch without changing Look-back Windows, and Batch Size. The standard deviation (STD) is used to evaluate the sensitivity of our framework. The formula for STD is as follows:

\begin{equation}
\text{\textit{STD}} = \sqrt{\frac{1}{n}\sum_{i=1}^{n}(M_i - \bar{M})^2}, 
\end{equation}
where we define $M_i$ as the evaluation metric observed for each hyperparameter adjustment, and $\bar{M}$ as the average of these $M_i$ values. It's important to note that changes to these three hyperparameters primarily occur during the large-sample data pre-training stage. Their values are not altered during subsequent transfer and fine-tuning. Moreover, the results are the predictions on the few-shot datasets. \textbf{Table 8} shows our experimental result.

\begin{table}[h]
\small
\begin{spacing}{1.4}
\caption{\textbf{Sensitive analysis result}}
\small
\setlength{\tabcolsep}{7.4mm}{
\begin{tabular}{ccccc}
\hline
\multirow{2}{*}{}              & \multirow{2}{*}{\textbf{Index}} & \multicolumn{3}{c}{\textbf{IDS-Net (Predictions on Small Datasets)}}       \\ \cline{3-5} &        & \textbf{$ME\Phi (LW,BS)$} & \textbf{$LW\Phi (ME,BS)$} & \textbf{$BS\Phi (LW,ME)$}  \\ \hline
\multirow{3}{*}{\textbf{Data 1}} & \textbf{RMSE}   & 0.0205    &0.0302    & 0.0096   \\
   & \textbf{MAE}   &  0.0560   &  0.0400    & 0.0230    \\
 & \textbf{$R^2$}  & 0.0009    &  0.0013    &   0.0004  \\ \hline
\multirow{3}{*}{\textbf{Data 2}} & \textbf{RMSE}   & 0.2358 & 0.1026  & 0.0655 \\
   & \textbf{MAE}   & 0.1419    &   0.0739   &  0.0567 \\
 & \textbf{$R^2$}   & 0.0061   &  0.0016    & 0.0009    \\ \hline
\end{tabular}}
\end{spacing}
\end{table}

Experimental results from the sensitivity analysis in \textbf{Table 8} demonstrate that the prediction performance of our proposed IDS-Net is minimally affected by hyperparameters, which demonstrates the excellent practical engineering applicability of our proposed model.

\vspace{0.25\baselineskip}

\textbf{Remark 4.} The study evaluated how well the model maintained its performance despite adjustments to hyperparameters. The outcomes clearly show that our proposed model operates largely independently of such variations, proving its strong capability for real-world engineering deployment.

\section{Conclusion}

We introduce a novel PV prediction framework that leverages interpretable adaptive selection and feature information fusion strategies to effectively solve the problem of limited historical training data for newly constructed PV power stations. Through four designed experiments, our proposed framework comprehensively validates its effectiveness, with prediction accuracy significantly surpassing both traditional deep learning models and the latest state-of-the-art time series forecasting models. Specifically, our framework is primarily divided into two main parts: a data preprocessing module and a PV power prediction module. In the data preprocessing module, we first utilize the MMD technique to select the source domain dataset with the most similar data distribution. Then, we proposed the HI algorithm for outlier processing. Finally, the ReliefF algorithm is used for feature selection to avoid feature redundancy. In the PV power prediction module, we first design a deep learning prediction model pool for PV forecasting. Then, a parallel dual-channel ensemble strategy is used for model interpretability and adaptive selection. Moreover, we construct a cross-correlation embedding strategy for feature information fusion. Finally, we design an MLP layer for the prediction results. In the transfer learning stage, we design an end-to-end transfer strategy for adaptive selection in the fine-tuning strategy. In the training process, we proposed a novel loss function with a penalty weighting strategy, which can make the model achieve more accuracy on peak values. Our proposed PV forecasting framework was validated on two PV power datasets, achieving an accuracy exceeding other rivals.

\vspace{0.25\baselineskip}
While our model achieves excellent prediction accuracy, there's still room for improvement in future research. First, we aim to integrate multi-modal information using large language model techniques to improve PV prediction accuracy. Second, we hope to extend our model's prediction horizon to longer sequences, improving its long-sequence forecasting ability.

\section*{Acknowledgment}

This research is supported by the General Program of the Natural Science Foundation of Hebei Province under the project "Multi-temporal and Spatial Scale Power Forecasting of Distributed Photovoltaics Based on Generative Artificial Intelligence" (Grant No. G2025502003).

\section*{Data availability}

Data will be made available on request.

\bibliographystyle{elsarticle-num} 
\bibliography{cas-refs}

\end{document}